\documentstyle[12pt,moriond,psfig]{article}
\def\be{\begin{equation}}

\def\ee{\end{equation}}

\def\etal{et al.}

\def\hMpc{h^{-1}{\rm Mpc}}

\def\hcMpcinv{h^{3}{\rm Mpc}^{-3}}

\newcommand{\Halpha}{\mbox{H$\alpha$}}
\newcommand{\Hi}{\mbox{H\,{\sc i}}}
\newcommand{\Hii}{\mbox{H\,{\sc ii}}}

\newcommand{\ew}{\mbox{EW}}
\newcommand{\ewh}{\mbox{EW\,(H$\alpha$)}}
\newcommand{\ewo}{\mbox{EW\,([O\,{\sc ii}])}}
\begin{document}
%Here you should enter the title of your manuscript
\heading{LUMINOSITY AND CLUSTERING OF GALAXIES SELECTED BY \ewh}

\author{J. Loveday $^{1}$, L. Tresse $^{2}$, S.J. Maddox $^{3}$} 
{$^{1}$ Astronomy and Astrophysics, University of Chicago, Chicago, USA.}  
{$^{2}$ Istituto di Radioastronomia del CNR, Bologna, Italia.} 
{$^{3}$ Institute of Astronomy, Cambridge, UK.} 

\begin{moriondabstract}
We study the luminosity function and clustering properties of subsamples
of local galaxies selected from the Stromlo-APM survey by the rest-frame
equivalent width (EW) of the \Halpha\ emission line.
The $b_J$ luminosity function of star-forming galaxies has a significantly 
steeper faint-end slope than that for quiescent galaxies:
the majority of sub-$L^*$ galaxies are currently undergoing significant
star formation.
Emission line galaxies are less strongly clustered, both amongst themselves,
and with the general galaxy population, than quiescent galaxies.
Thus as well as being less luminous, star-forming galaxies also inhabit
lower-density regions of the Universe than quiescent galaxies.
\end{moriondabstract}

\section{Galaxy Samples} \label{sec:samples}

Our sample of galaxies is taken from the Stromlo-APM redshift
survey which covers 4300 sq-deg of the south galactic cap and consists
of 1797 galaxies brighter than $b_J = 17.15$ mag.  The galaxies all
have redshifts $z < 0.145$, and the mean is $\langle z \rangle =
0.051$.  A detailed description of the spectroscopic observations and
the redshift catalog is published in \cite{lpme96}.
Measurement of \ewh\ and other spectral properties is described in
\cite{tmls99} and a more detailed analysis of the luminosity function and
clustering of galaxies selected by \ewh\ as well as \ewo\ may be found in
\cite{ltm99}.
Of the 1797 galaxies originally published in the redshift survey, 
1521 are suitable for analysis here.
The rest are brighter than 15th magnitude, and so have unreliable APM
photometry, or have a problem with the spectrum meaning that \ewh\ could
not be accurately measured.

We select galaxy subsamples using measured equivalent width of the \Halpha\
emission line, the most reliable tracer of massive star formation 
\cite{kenn83}.
The \Halpha\ line is detected with \ew\ $\ge 2$\AA\ in 61\% of galaxies.
Of these emission-line galaxies, half have \ewh\ $ > 15$\AA.
Thus we form three subsamples of comparable size by dividing the sample at
\ewh\ of 2\AA\ and 15\AA.
The galaxy samples selected by \Halpha\ equivalent width are defined
in Table~\ref{tab:samples} which also gives the numbers of galaxies of each
morphological type in each spectroscopically selected subsample.
The sample labeled ``Unk'' consists of galaxies to which no morphological
classification was assigned.
We see that early-type galaxies dominate when \Halpha\
emission is not detected and are underrepresented when emission is detected.
Conversely, the number of irregular galaxies increases significantly in the 
spectroscopic samples which show strongest star formation.

\begin{table}
 \begin{center}
 \caption{Spectroscopic subsamples and correlation with morphological type.}
 \vspace{1mm}
 \label{tab:samples}
 \begin{tabular}{llrrrrrr}
 \hline
 \hline
 Sample & \ewh & E & S0 & Sp & Irr & Unk & Total\\
 \hline
 (a) H-low & $ < 2$ \AA & 125 & 108 & 207 & 10 & 149 & 599\\
 (b) H-mid & 2--15 \AA &   8 &  16 & 340 & 18 &  81 & 463\\
 (c) H-high & $ > 15$ \AA &  11 &   9 & 303 & 41 &  95 & 459\\
  \hline
  \hline
 \end{tabular}
\end{center}
\end{table}

\vspace{-5mm}
\section{The Galaxy Luminosity Function} \label{sec:LF}

% lum:lumfig_spec.csh, lum:lumfig.f, lumfn_ha*.dat, phi_ha*.dat, like_ha*.dat
\begin{figure*}[htbp]
 \vspace{-15mm}
\centerline{\psfig{figure=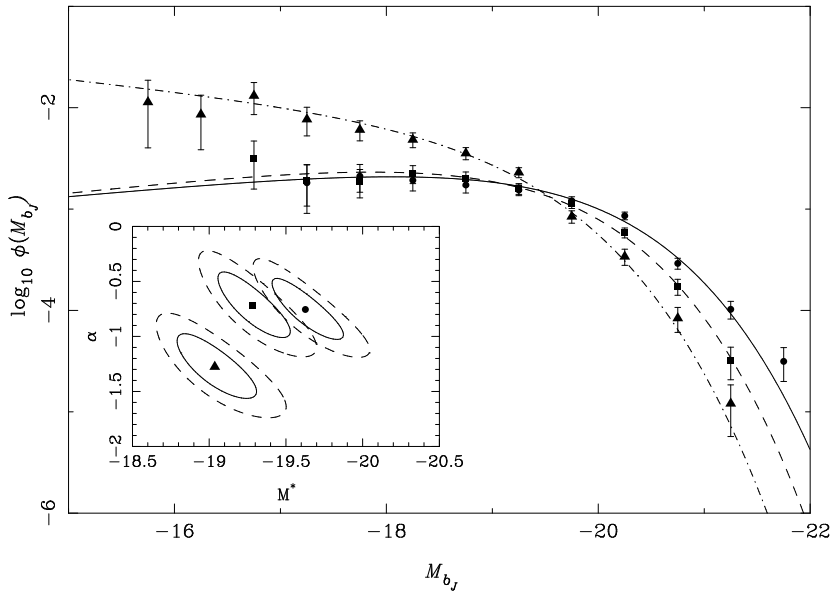,width=120mm,angle=-90}}
\caption{\setlength{\baselineskip}{0.3cm}
 Estimates of the luminosity function for galaxies with no
significant detected \Halpha\ emission (H-low: circles, solid line),
with moderate \Halpha\ emission (H-mid: squares, dashed
line) and with strong \Halpha\ emission (H-high: triangles, dot-dashed line).
The symbols with error bars show a stepwise fit, the curves show
Schechter function fits.
For clarity, data points representing fewer than five galaxies have been
omitted from the plot.
The inset shows 1 \& 2$\sigma$ likelihood contours for the best-fit 
Schechter parameters.
\label{fig:lf_ha}}
\vspace{-2mm}
\end{figure*}

Our estimates of the luminosity function for the \ewh\ selected samples,
assuming a Hubble constant of $H_0 = 100$ km/s/Mpc,
are shown in Figure~\ref{fig:lf_ha}.
The inset to this Figure shows the likelihood contours for the best-fit 
Schechter parameters $\alpha$ and $M^*$.
The Schechter parameters and their $1\sigma$ errors (from the bounding box
of the $1\sigma$ error contours) are also listed in Table~\ref{tab:LF}.
Note that the estimates of $\alpha$ and $M^*$ are strongly correlated
and so the errors quoted for $\alpha$ and $M^*$ in the Table are conservatively
large.
We see a trend of faintening $M^*$ and steepening $\alpha$ as \ewh\ increases.

\begin{table*}[htbp]
 \begin{center}
 \caption{Luminosity and correlation function parameters.}
 \vspace{1mm}
 \label{tab:LF}
 \label{tab:corr}
 \begin{math}
 \begin{array}{lcccccc}
 \hline
 \hline
 {\rm Sample} & \alpha & M^* & \phi^* & \rho_L & \gamma & r_0\\
 \hline
 \mbox{(a) H-low} & -0.75 \pm 0.28 & -19.63 \pm 0.24 & 
4.5 \pm 1.1 & 5.9 \pm 1.4 & 1.78 \pm 0.08 & 6.0 \pm 1.4\\
 \mbox{(b) H-mid} & -0.72 \pm 0.29 & -19.28 \pm 0.23 & 
5.4 \pm 1.4 & 5.1 \pm 1.4 & 1.60 \pm 0.13 & 5.2 \pm 2.0\\
 \mbox{(c) H-high} & -1.28 \pm 0.30 & -19.04 \pm 0.26 & 
8.5 \pm 2.8 & 8.8 \pm 2.9 & 1.87 \pm 0.16 & 2.9 \pm 1.9\\
  \hline
  \hline
 \end{array}
 \end{math}
 \medskip

\setlength{\baselineskip}{0.3cm}
Notes: $\alpha$ is the faint-end slope and $M^*$ the characteristic $b_J$
magnitude of the best-fit Schechter function.
$\phi^*$ is the normalisation of the Schechter luminosity function,
in units of $10^{-3} \hcMpcinv$.
$\rho_L$ is the luminosity density in the range $-22 < M < -15$, 
in units of $10^7 L_{\odot}\hcMpcinv$.
$\gamma$ and $r_0$ are the best-fit power-law parameters to the correlation 
function.
\end{center}
\end{table*}

The normalisation $\phi^*$ of the fitted Schechter functions was estimated
using a minimum variance estimate of the space density $\bar{n}$ of galaxies
in each sample \cite{dh82}, \cite{lpem92}.
We corrected our estimates of $\bar{n}$, $\phi^*$ and luminosity density 
$\rho_L$ to allow for those galaxies excluded from each subsample.
The uncertainty in mean density due to ``cosmic variance''
is $\approx 6\%$ for each sample.
However, the errors in these quantities are dominated by the uncertainty
in the shape of the LF, particularly by the estimated value of the 
characteristic magnitude $M^*$.

Using \Halpha\ equivalent width as an indicator of star formation
activity, we find that galaxies currently undergoing significant bursts of 
star formation
dominate the faint-end of the luminosity function, whereas more quiescent
galaxies dominate at the bright end.
This is in agreement with the results from the LCRS \cite{lin96a} and ESP
\cite{zucc97} surveys for samples selected by \ewo.

\section{Galaxy Clustering} \label{sec:clust}

\begin{figure}[htbp]
\centerline{\psfig{figure=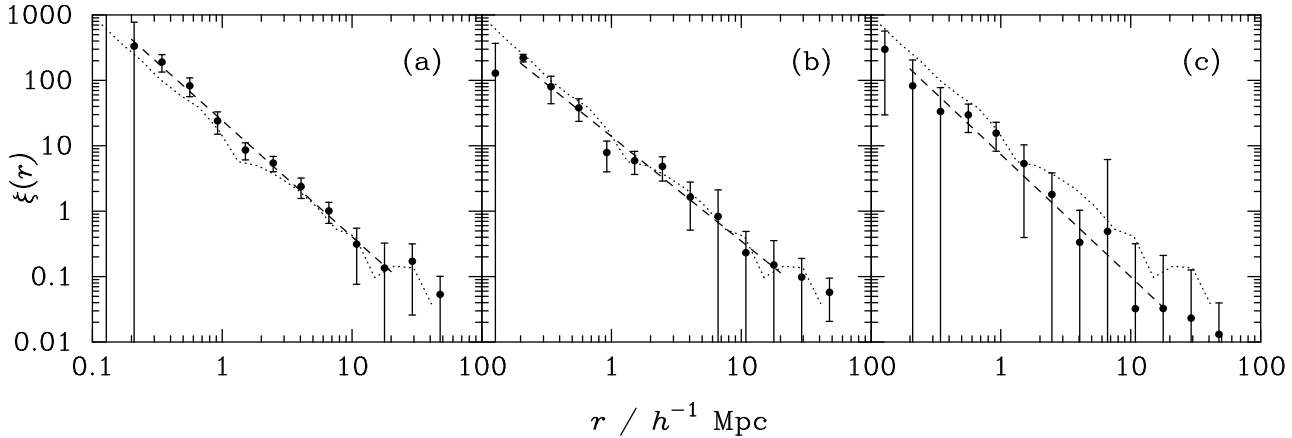,width=170mm,angle=-90}}
\caption{\setlength{\baselineskip}{0.3cm}
Estimates of the real-space correlation function for the
galaxy samples given in Table~\protect{\ref{tab:samples}}.
Error bars show the rms variance from dividing the survey into 4 distinct 
zones.
The dashed line shows the best-fit power-law over the range 0.2--20 $\hMpc$.
The dotted line shows $\xi(r)$ estimated from the full Stromlo-APM sample
\cite{lmep95}.
\label{fig:xir}}
\end{figure}

We have calculated
the projected cross-correlation function $\Xi(\sigma)$ of each galaxy subsample
with all galaxies in the APM survey to a magnitude limit of $b_J = 17.15$.
We then invert this projected correlation function to obtain the real space
cross-correlation function $\xi(r)$ of each subsample with the full galaxy 
sample.
This method of estimating $\xi(r)$ is described by \cite{srl92}
and by \cite{lmep95}.
Our estimates of $\xi(r)$ are plotted in Figure~\ref{fig:xir} and our best-fit
power-laws are tabulated in Table~\ref{tab:corr}.
We see that strong emission-line galaxies are more 
weakly clustered than their quiescent counterparts by a factor of about two.

% corr:xip_mult.f: xip_xcorr_all.dat, xcp_hlow/mid/high.dat, 

The clustering measured for non-ELGs is very close to that measured 
for early-type (E + S0) galaxies, and the clustering of
late-type (Sp + Irr) galaxies lies between that of the
moderate and high EW galaxies (cf. \cite{lmep95}).
Given the strong correlation between morphological type and presence of
emission lines (Table~\ref{tab:samples}) this result is not unexpected.
The power-law slopes are consistent ($\gamma_r = 1.8 \pm 0.1$) 
between the H-low and H-high samples.
For the H-mid sample
we find shallower slopes ($\gamma_r = 1.6 \pm 0.1$).
This is only a marginally significant (1--2 $\sigma$) effect,
but may indicate a deficit of moderately star-forming galaxies principally
in the cores of high density regions, whereas strongly star forming
galaxies appear to more generally avoid overdense regions.

\section{Conclusions} \label{sec:concs}

We have presented the first analysis of the luminosity function and spatial
clustering for representative and well-defined local
samples of galaxies selected by 
\ewh, the most direct tracer of star-formation.
We observe that $M^*$ faintens systematically with increasing \ewh, and that
the faint-end slope increases.
Star-forming galaxies are thus likely to be significantly fainter than their 
quiescent counterparts.
The faint-end ($M \gsim M^*$) of the luminosity function is dominated by ELGs
and thus the majority of local dwarf galaxies are currently undergoing 
star formation.
Star-forming galaxies are more weakly clustered than quiescent galaxies.
This weaker clustering is observable on scales from 0.1--10 $\hMpc$.
We thus confirm that star-forming galaxies are preferentially found today in 
low-density environments.

A possible explanation for these observations is that
luminous galaxies in high-density 
regions have already formed all their stars by today, while less luminous 
galaxies in low-density regions are still undergoing star formation.
It is not clear what might be triggering the star formation in these galaxies
today.
While interactions certainly enhance the rate of star formation in some
disk galaxies, interactions with luminous companions can only
account for a small fraction of the total star formation in disk
galaxies today (\cite{kkhhr87}).
Telles \& Maddox \cite{tm99} have investigated the environments of \Hii\
galaxies by cross-correlating a sample of \Hii\
galaxies with APM galaxies as faint as $b_J = 20.5$.
They find no excess of companions with \Hi\ mass $\gsim 10^8 M_{\odot}$ near
\Hii\ galaxies, thus arguing that star formation in most \Hii\ galaxies is
unlikely to be induced by even a low-mass companion.

Our results are entirely consistent with the hierarchical picture of
galaxy formation.
In this picture, today's luminous spheroidal galaxies formed from
past mergers of galactic sub-units in high density regions, and produced all
of their stars in a merger induced burst, or series of bursts, over a 
relatively short timescale.
The majority of present-day dwarf, star-forming galaxies in lower density
regions may correspond to unmerged systems formed at lower peaks in the
primordial density field
and whose star formation is still taking place.
Of course, the full picture of galaxy formation is likely to be significantly
more complicated than this simple sketch, and numerous physical effects
such as depletion of star-forming material and other feedback mechanisms
are likely to play an important role.

% References listed in alphabetical order ...

%
\vfill
\end{document}